\documentclass[conference,a4paper]{IEEEtran}
 \usepackage{amsmath,amssymb,graphicx,booktabs,multirow}
\usepackage{cite}
\usepackage{xcolor}
\usepackage{url}

 \usepackage{pgfplots}
 \usepgfplotslibrary{groupplots}
 \usetikzlibrary{calc}
 \pgfplotsset{compat=1.18}

\newcommand{\vx}{\mathbf{x}}
\newcommand{\vy}{\mathbf{y}}
\newcommand{\vmu}{\boldsymbol{\mu}}
\newcommand{\mH}{\mathbf{H}}
\newcommand{\mP}{\mathbf{P}}
\newcommand{\mS}{\boldsymbol{\Sigma}}
\newcommand{\mB}{\mathbf{B}}

\newcommand{\mW}{\mathbf{W}}
\newcommand{\E}{\mathbb{E}}
\newcommand{\argmax}{\operatorname*{arg\,max}}
\newcommand{\softmax}{\operatorname{softmax}}
\newcommand{\vn}{\mathbf{n}}

\newcommand{\mG}{\mathbf{G}}
 \newcommand{\mI}{\mathbf{I}}
\newcommand{\vh}{\mathbf{h}}
 \newcommand{\vdelta}{\boldsymbol{\delta}}

\title{Classification-oriented adaptive sensing via posterior sampling}

\author{\IEEEauthorblockN{Andriy Enttsel, Maxime Rousselot, Vincent Corlay}
\IEEEauthorblockA{Mitsubishi Electric R\&D Centre Europe}
}

\begin{document}
\maketitle
\begin{abstract} 
Recent advances in diffusion models have enabled high-performance, instance-adaptive compressed sensing through posterior sampling, without task-specific policy training. Existing methods select sensing probes by maximizing total posterior signal variance and are therefore primarily reconstruction-driven. We introduce a classification-driven extension motivated by the closed-form posterior covariance of a class-conditional Gaussian mixture model, which decomposes into within-class and between-class uncertainty. Using calibrated soft classifier outputs, we estimate these uncertainty terms from diffusion posterior samples and propose a classification-oriented criterion for selecting the dominant sensing direction in the unmeasured subspace. Experiments on MNIST and CIFAR-10 compare the resulting classification accuracy, measurement cost, and reconstruction quality with those of reconstruction-oriented counterparts. The results identify regimes in which semantic posterior uncertainty yields a more favorable classification--measurement trade-off and quantify the associated reconstruction cost. 
\end{abstract}


\section{Introduction}
\label{sec:intro}

Compressed sensing recovers sparse signals from fewer linear measurements than their ambient dimension~\cite{candes2006robust,donoho2006compressed}. In classical compressed sensing, the sensing operator is fixed before observing the unknown signal and is often drawn from a universal random ensemble. 

More informed sensing designs can improve performance by exploiting statistical or physical prior knowledge, by adapting sequentially to the observed signal, or by targeting a downstream task rather than reconstruction alone.

Rakeness-based sensing, for example, uses population-level second-order signal statistics to increase the captured signal energy while preserving measurement diversity~\cite{mangia2012rakeness}. Physical knowledge can also guide measurement design, as demonstrated in vibration-based structural health monitoring using modal information and signal-adapted representations ~\cite{zonzini2021model,ravaglia2026adaptive}. In both cases, the sensing operator exploits a signal model but is generally not adapted to the measurements of an individual test instance. 

%Task-oriented compressed sensing instead designs measurements for inference. Detection, classification, estimation, and filtering can be performed directly from compressive measurements without first reconstructing the signal~\cite{davenport2010signal}. The smashed filter, for example, demonstrated compressed-domain target classification using a single-pixel camera~\cite{davenport2007smashed}. These results show that measurements suited to reconstruction need not be optimal for classification. However, many task-oriented methods design the sensing operator in advance rather than adapting it to the uncertainty remaining for each observed signal. 

Task-oriented compressed sensing instead designs the sensing operator for inference, enabling detection, classification, estimation, or filtering directly from compressive observations~\cite{davenport2010signal}. The smashed filter, for example, demonstrated this principle for compressed-domain target classification with a single-pixel camera~\cite{davenport2007smashed}. More generally, these methods show that measurements optimized for reconstruction may be suboptimal for classification, but they typically rely on a fixed sensing operator rather than adapting the acquisition to the remaining observation-specific uncertainty.

Gaussian-mixture models (GMMs) provide a tractable framework combining statistical prior information, task-oriented measurement design, and sequential adaptation. Measurements can be selected to discriminate among plausible mixture components before optional component-conditioned reconstruction~\cite{yu2011statistical, carson2012communications, duarte2013task, braun2015infogreedy}. However, GMMs may be too restrictive for complex signal distributions.
%
%Deep neural networks and diffusion models replace restrictive parametric priors with richer learned representations of the signal distribution~\cite{ho2020ddpm}. Incremental schemes with learned performance prediction have been proposed for MRI and ECG signals~\cite{marchioni2023i2mtc,martinini2025sp}, adapting the acquisition length to each instance but not the sensing directions to the remaining uncertainty. 
%For sequentially adaptive image sensing, AdaSense~\cite{elata2024adasense} is a diffusion-based, training-free approach that samples from the posterior conditioned on the measurements acquired so far, selects the unmeasured direction of maximum posterior variance, and resamples after each acquisition.
%The framework has also been extended to lossy compression~\cite{elata2025psc} and semantic communication~\cite{bingxuan2026semcom}. Nevertheless, maximizing total posterior variance remains primarily reconstruction-oriented and may prioritize uncertainty of factors weakly related to the remaining class ambiguity. 

Deep neural networks and diffusion models replace restrictive parametric priors with richer learned signal models~\cite{schwab2019ip, ho2020ddpm}. Incremental schemes with learned performance prediction have been proposed for MRI and ECG~\cite{marchioni2023i2mtc,martinini2025sp}, adapting the acquisition length to each instance but not the sensing directions to the remaining uncertainty. AdaSense~\cite{elata2024adasense} instead performs sequentially adaptive image sensing by sampling from the measurement-conditioned diffusion posterior, selecting the unmeasured direction of maximum posterior variance, and resampling after each acquisition. The framework has also been extended to lossy compression~\cite{elata2025psc} and semantic communication~\cite{bingxuan2026semcom}. However, maximizing total posterior variance remains reconstruction-oriented and may prioritize sources of variability weakly related to the remaining class ambiguity.

Motivated by this limitation, we introduce a classification-oriented extension of AdaSense that directs sensing toward uncertainty relevant to the remaining class decision. A pretrained classifier assigns soft class probabilities to diffusion posterior samples, allowing us to estimate a sample-based analogue of the between-class covariance from the GMM analysis. The next probe is selected along the dominant direction of this semantic covariance after projection onto the unmeasured subspace. After each acquisition, the posterior is resampled and the semantic covariance recomputed, allowing the sensing direction to adapt to evolving class ambiguity. The same posterior-predictive probabilities define a confidence-based stopping rule, yielding an image-dependent measurement budget.

Our contributions are threefold: \textit{i}) a GMM analysis of AdaSense showing why maximizing total posterior variance can be suboptimal for classification; \textit{ii}) a classification-oriented specialization that uses soft class assignments to estimate semantic between-class covariance from diffusion posterior samples and combines adaptive probe design with confidence-based stopping; and \textit{iii}) an evaluation on MNIST and CIFAR-10 that characterizes the trade-off among classification accuracy, measurement cost, and reconstruction quality.

\section{Background: Reconstruction-Oriented Adaptive Sensing with Posterior Sampling}
\label{sec:background}

% Let $\vx\in\mathbb{R}^{D}$ denote an unknown signal, and let
% \begin{equation}
%     \vy_t = \mH_t\vx + \mathbf{n}_t, \qquad \mathbf{n}_t \sim \mathcal{N}(0, \sigma_n^2 \mathbf{I})
%     \label{eq:measurement_model}
% \end{equation}
% collect the $m_t$ linear measurements available at stage $t$. The rows of
% $\mH_t$ are assumed to be orthonormal. At each stage, a unit-norm probe
% $\vh_{t+1}$ is selected from the current information state
% $(\mH_t,\vy_t)$. The corresponding scalar measurement is acquired, and
% $\vh_{t+1}^{\top}$ is appended as a new row of $\mH_t$.

Let $\vx\in\mathbb{R}^{D}$ denote an unknown signal. At stage $t$, let
\begin{equation}
    \vy_t = \mH_t\vx + \mathbf{n}_t,
    \qquad
    \mathbf{n}_t \sim \mathcal{N}\!\left(\mathbf{0}, \sigma_n^2 \mathbf{I}_t\right),
    \label{eq:measurement_model}
\end{equation}
where $\vy_t\in\mathbb{R}^t$ collects the $t$ linear measurements acquired
so far, and $\mH_t\in\mathbb{R}^{t\times D}$ stacks the corresponding
sensing probes as rows. The probes are chosen to be orthonormal, i.e.,
$\mH_t\mH_t^\top=\mathbf{I}_t$.

At each stage, a unit-norm probe $\vh_{t+1}\in\mathbb{R}^D$ is selected
from the current information state $(\mH_t,\vy_t)$. The corresponding
scalar measurement is then acquired, and $\vh_{t+1}^{\top}$ is appended
as a new row of $\mH_t$.

\begin{figure*}[t]
\centering
\includegraphics[width=0.8\textwidth]{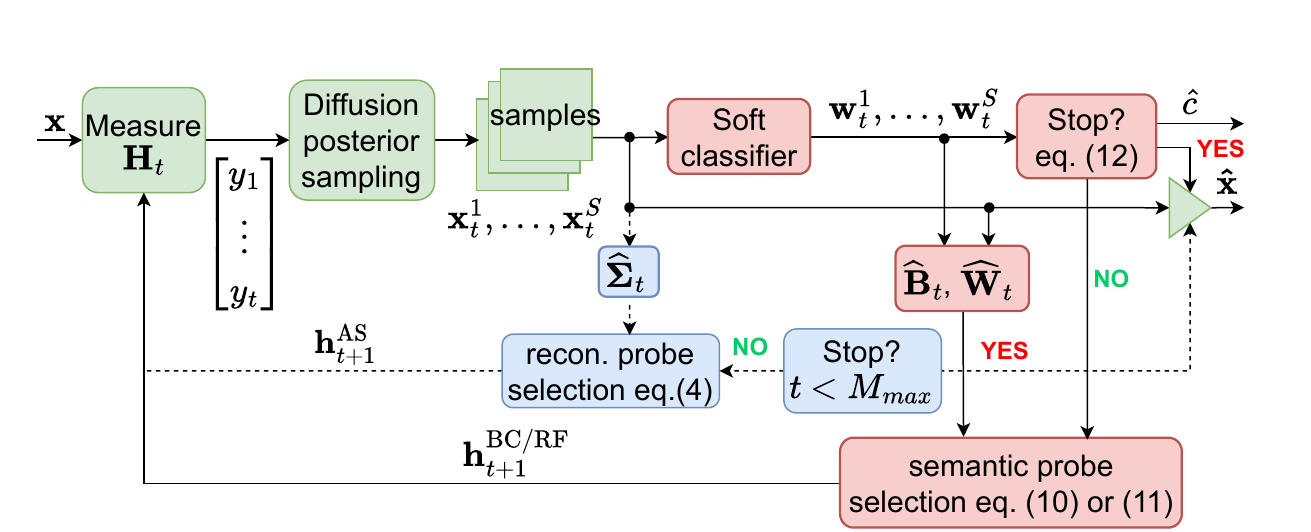}
\caption{Posterior sampling-based adaptive sensing for reconstruction (blue) and classification (red). The shared blocks are in green.}
\label{fig:scheme}
\end{figure*}

%In AdaSense~\cite{elata2024adasense}, a diffusion inverse solver generates samples from the
%measurement-conditioned posterior,
%\begin{equation}
%    \vx_t^{(s)}
%    \sim q_\theta(\vx\mid\mH_t,\vy_t),
%    \qquad s=1,\ldots,S.
%    \label{eq:posterior_samples}
%\end{equation}
In AdaSense~\cite{elata2024adasense}, a diffusion-based inverse solver generates samples from the measurement-conditioned posterior, 
\begin{equation} 
\vx_t^{(s)} \sim q_\theta\!\left(\vx \mid \mH_t,\vy_t\right), \quad s=1,\ldots,S.
 \label{eq:posterior_samples} 
\end{equation}
where the samples are constrained to be consistent with the acquired measurements up to the assumed measurement noise, that is, $ \|\mH_t\vx_t^{(s)}-\vy_t \|_2 \leq \epsilon_t$ \cite{kawar2022ddrm} . 
The empirical sample posterior mean and unbiased covariance then are
\begin{align}
    \widehat{\vmu}_t
    &= \frac{1}{S}\sum_{s=1}^{S}\vx_t^{(s)},
\quad
    \widehat{\mS}_t
    = \frac{1}{S-1}\sum_{s=1}^{S} \widetilde{\vx}_t^{(s)} \bigl(\widetilde{\vx}_t^{(s)}\bigr)^\top
    \label{eq:posterior}
\end{align}
where $\widetilde{\vx}_t^{(s)}=\vx_t^{(s)}-\widehat{\vmu}_t$.
AdaSense selects the next probe as the maximum-variance direction in the unmeasured subspace:
\begin{equation}
    \vh_{t+1}^{\mathrm{AS}}
    =
    \argmax_{\substack{\|\vh\|_2=1, \mH_t\vh=\mathbf{0}}}
    \vh^\top\widehat{\mS}_t\vh.
    \label{eq:adasense_optimization}
\end{equation}
The constraint $\mH_t\vh=\mathbf{0}$ prevents previously measured directions
from being selected again. Since the rows of $\mH_t$ are orthonormal, $ \mP_t = \mI-\mH_t^\top\mH_t$
is the orthogonal projector onto the residual sensing subspace. Hence,
\eqref{eq:adasense_optimization} is solved by a dominant eigenvector of
$\mP_t\widehat{\mS}_t\mP_t$, corresponding to the direction of largest
remaining posterior image uncertainty\footnote{To be precise, the original AdaSense formulation assumes $\epsilon_t=0$ and strict measurement consistency, $\mH_t\vx_t^{(s)}=\vy_t$, so that the next probe is obtained directly as $\vh_{t+1}=\operatorname{eigvec}_{\max}(\widehat{\mS}_t)$. For low-noise measurements or approximate measurement consistency, we instead introduce $\mP_t$ to remove residual variance along previously measured directions.}.
After each
acquisition, the diffusion posterior is resampled using the updated state
$(\mH_{t+1},\vy_{t+1})$. In the standard reconstruction-oriented setting,
this process continues until the prescribed maximum measurement budget $M_{\max}$ is
reached and the final posterior mean $\widehat{\vmu}_{M_{\max}}$ is the MMSE estimated reconstruction of the signal. This approach is summarized in Fig.~\ref{fig:scheme} (blue).

\section{Analytical Specialization with a Gaussian Mixture Posterior}
\label{sec:gmm_specialization}

To expose the distinction between reconstruction and classification
uncertainty, we study AdaSense under a tractable GMM prior,
\begin{equation}
    p(\vx)
    = \sum_{c=1}^{K}w_c\,
    \mathcal{N}(\vx;\vmu_c,\mS_c),
    \qquad
    w_c\geq 0,
    \quad
    \sum_{c=1}^{K}w_c=1,
    \label{eq:gmm_prior}
\end{equation}
where $w_c$, $\vmu_c$, and $\mS_c$ denote, respectively, the mixing weight, mean vector, and covariance matrix of component $c$.
Assuming Gaussian measurement noise with covariance $\mathbf{R}_t$ we can define the measurement covariance $\mG_{c,t} = \mH_t\mS_c\mH_t^\top+\mathbf{R}_t$.
Conditioning each component on $\vy_t$ gives
\begin{align}
    w_{c,t}
    &=
    \frac{w_c\,
    \mathcal{N}(\vy_t;\mH_t\vmu_c,\mG_{c,t})}
    {\sum_{j=1}^{K}w_j\,
    \mathcal{N}(\vy_t;\mH_t\vmu_j,\mG_{j,t})},
    \label{eq:resp_main}\\
    \vmu_{c\mid t}
    &= \vmu_c
    + \mS_c\mH_t^\top\mG_{c,t}^{-1}
    (\vy_t-\mH_t\vmu_c),
    \label{eq:cond_mean_main}\\
    \mS_{c\mid t}
    &= \mS_c
    - \mS_c\mH_t^\top\mG_{c,t}^{-1}
    \mH_t\mS_c.
    \label{eq:cond_cov_main}
\end{align}
The posterior mean is $\overline{\vmu}_t = \sum_{c=1}^{K}w_{c,t}\vmu_{c\mid t}.$
By the law of total covariance, the posterior covariance decomposes as
\begin{align}
    \mS_t^{\mathrm{post}}
    &=
    \underbrace{\sum_{c=1}^{K}w_{c,t}\mS_{c\mid t}}_{\mW_t}
    +
    \underbrace{\sum_{c=1}^{K}w_{c,t}
    (\vmu_{c\mid t}-\overline{\vmu}_t)
    (\vmu_{c\mid t}-\overline{\vmu}_t)^\top}_{\mB_t}.
    \label{eq:gmm_decomp}
\end{align}
Proof of \eqref{eq:resp_main}-\eqref{eq:cond_cov_main} is provided in the Appendix. The pair $(\overline{\vmu}_t,\mS_t^{\mathrm{post}})$ is the closed-form
counterpart of the sample-based quantities in~\eqref{eq:posterior}. Replacing
$\widehat{\mS}_t$ in \eqref{eq:adasense_optimization} by
$\mS_t^{\mathrm{post}}$ therefore gives a sampling-free GMM specialization of
reconstruction-oriented posterior-covariance sensing. The posterior mean
$\overline{\vmu}_t$ is the corresponding MMSE estimator.

When mixture components represent semantic classes\footnote{For reconstruction-only settings, the GMM can be fitted without labels.}, the decomposition in
\eqref{eq:gmm_decomp} separates two distinct sources of uncertainty. The within-class covariance $\mW_t$ captures the residual signal variability conditioned on the class, including intrinsic class variability, nuisance factors, and reconstruction details. In contrast, the between-class covariance $\mB_t$
measures disagreement among the current class-conditioned
posterior means. Consequently, the mode ambiguity is already addressed implicitely in a soft manner, but maximizing
$\vh^\top(\mW_t+\mB_t)\vh$ can still favor directions with substantial signal variance but limited discriminative value.

%Adaptive sensing with GMM priors has also been studied using
%component-specific and information-theoretic criteria.
%Duarte-Carvajalino et al.~\cite{duarte2013task} first identify a Gaussian
%component and subsequently adapt reconstruction measurements to its covariance
%and the previously acquired observations. Carson et al.~\cite{carson2012communications} 
%consider full-distribution information-theoretic designs and a fast online
%approximation based on the dominant posterior component. Braun et
%al.~\cite{braun2015infogreedy} develop conditional information-greedy sensing
%and, in the low-SNR regime, use the posterior-weighted within-component
%covariance $\mW_t$. Our use of the GMM is complementary: it provides an analytical
%specialization of posterior-covariance sensing and makes the between-class term
%$\mB_t$ explicit.

It is worth noting that, prior GMM-based methods exploit different posterior quantities: Duarte-Carvajalino et al.~\cite{duarte2013task} use the covariance of an identified component $\mS_{c\mid t}$, Carson et al.~\cite{carson2012communications} use information-theoretic criteria over the mixture or its dominant component $\mS_{c\mid t}$, and Braun et al.~\cite{braun2015infogreedy} use the posterior-weighted within-component covariance $\mW_t$ at low SNR. Our use of the GMM is complementary: it provides an analytical specialization of posterior-covariance sensing and highlights the importance of the between-class term $\mB_t$ also for reconstruction.

\definecolor{AdaBlue}{HTML}{3a86ff} %0072B2
\definecolor{BetweenOrange}{HTML}{ff006e} %D55E00 
\definecolor{FisherGreen}{HTML}{57cc99} %009E73
\definecolor{PCAColor}{HTML}{c9ada7} %   242423
\definecolor{RandomColor}{HTML}{ff5400}  %e63946
\definecolor{AxisGray}{HTML}{4D4D4D}
\definecolor{GridGray}{HTML}{DEDEDE}
\definecolor{InsetGray}{HTML}{737373}
\definecolor{Black}{HTML}{000000}

\subsection{Classification-Oriented GMM Acquisition}
\label{subsec:gmm_classification}

The decomposition in \eqref{eq:gmm_decomp} separates the posterior
uncertainty remaining within each class from that induced by ambiguity among
classes. This suggests replacing the total covariance used by
reconstruction-oriented GMM sensing with its class-discriminative component.
The between-class (BC) rule is
\begin{equation}
    \vh_{t+1}^{\mathrm{BC}}
    =\argmax_{\substack{\|\vh\|_2=1, \mH_t\vh=\mathbf0}}
    \vh^\top\mB_t\vh,
    \label{eq:gmm_between_probe}
\end{equation}
whose solution is a dominant eigenvector of $\mP_t\mB_t\mP_t$. It therefore
targets the unobserved direction along which the plausible class-conditioned
posterior means disagree most.

The within-class variability can be reintroduced as a normalization that penalizes directions with high class-conditioned uncertainty. This yields the LDA-inspired regularized Fisher (RF) criterion~\cite{fisher1936use}
\begin{equation}
    \vh_{t+1}^{\mathrm{RF}}
    =\argmax_{\substack{\|\vh\|_2=1, \mH_t\vh=\mathbf0}}
    \frac{\vh^\top\mB_t\vh}
         {\vh^\top \left[\mW_t+\left(\gamma + \sigma_n^2 \right)\mI \right]\vh},
    \quad \gamma>0.
    \label{eq:gmm_fisher_probe}
\end{equation}
which can be solved as a generalized eigenvalue problem in the current residual subspace.
Thus, the three criteria differ only in the uncertainty they prioritize:
reconstruction-oriented sensing uses $\mW_t+\mB_t$,
\eqref{eq:gmm_between_probe} isolates $\mB_t$, and
\eqref{eq:gmm_fisher_probe} normalizes between-class disagreement by
within-class variability.

%The GMM responsibilities also define a classification confidence and a
%signal-dependent sensing budget. Under a correctly specified GMM,
%$w_{c,t}=p(c\mid\vy_t)$ is the exact posterior class probability. Hence,
%$\max_c w_{c,t}$ is the conditional probability that the MAP prediction is
%correct. Its expectation over $\vy_t$ equals the MAP classification accuracy $\mathbb{E}_{\vy_t}\!\left[\max_c w_{c,t}\right] = \Pr\!\left(c=\widehat{c}_t\right)$. This connection provides a probabilistic basis for the following signal-dependent stopping rule:
%\begin{align}
%    t^\star
%    &=\min\!\left\{t:
%      \max_c w_{c,t}\geq\tau_{\mathrm{stop}}
%      \ \text{or}\ m_t\geq M_{\max}\right\},
%    \\
%    \widehat c&=\argmax_c w_{c,t^\star}.
%    \label{eq:gmm_stopping_and_prediction}
%\end{align}
The GMM responsibilities also provide a classification-confidence measure and enable a signal-dependent sensing budget. Under a correctly specified GMM, $w_{c,t}=p(C=c\mid\vy_t)$ is the exact posterior probability of class $c$. For the MAP decision $\widehat{c}_t=\argmax_c w_{c,t}$, it follows that $\Pr(C=\widehat{c}_t\mid\vy_t)=\max_c w_{c,t}$. Thus, the maximum responsibility is the conditional probability that the MAP prediction is correct. Averaging over the measurement distribution yields the MAP classification accuracy, $\mathbb{E}_{\vy_t}[\max_c w_{c,t}]=\Pr(C=\widehat{C}_t)$. This identity provides a probabilistic basis for the following signal-dependent stopping rule and final prediction:
\begin{align}
 t^\star &= \min\!\left\{ t: \max_c w_{c,t}\geq\tau_{\mathrm{stop}} \ \text{or}\ m_t\geq M_{\max} \right\}, \label{eq:gmm_stopping}\\ \widehat{c} &= \argmax_c w_{c,t^\star}. \label{eq:gmm_stopping_and_prediction}
\end{align}

The threshold $\tau_{\mathrm{stop}}$ specifies the desired posterior
confidence, while $M_{\max}$ guaranties termination. 
%Equations
%\eqref{eq:gmm_between_probe}--\eqref{eq:gmm_stopping_and_prediction} define
%the acquisition, stopping, and prediction rules.
% The diffusion method below
%only estimates the GMM quantities appearing in these rules.

% ICASSP figure include: MNIST and CIFAR-10 sensing tradeoffs
% Usage: \input{improved_tradeoff_figure.tex}
%
% Required once in the manuscript preamble:
%   \usepackage{xcolor}
%   \usepackage{pgfplots}
%   \usepgfplotslibrary{groupplots}
%   \usetikzlibrary{calc}
%   \pgfplotsset{compat=1.18}

\pgfplotsset{
tradeoff axis/.style={
    width=7.15cm,
    height=3.55cm,
    scale only axis,
    grid=major,
    major grid style={draw=GridGray,line width=0.22pt},
    axis line style={draw=AxisGray,line width=0.42pt},
    tick style={draw=AxisGray,line width=0.35pt},
    tick label style={font=\footnotesize,color=AxisGray},
    label style={font=\small},
    title style={font=\small\bfseries,yshift=-1pt},
    every axis plot/.append style={mark=none},
    unbounded coords=discard,
    clip=true
},
tradeoff inset/.style={
    width=3.55cm,
    height=1.80cm,
    scale only axis,
    grid=major,
    major grid style={draw=GridGray,line width=0.18pt},
    axis background/.style={fill=white},
    axis line style={draw=AxisGray,line width=0.40pt},
    tick style={draw=AxisGray,line width=0.30pt},
    tick label style={font=\tiny,color=AxisGray},
    every axis plot/.append style={mark=none},
    unbounded coords=discard,
    clip=true
},
method adasense/.style={color=AdaBlue},
method between/.style={color=BetweenOrange},
method fisher/.style={color=FisherGreen},
method fisher2/.style={color=Black},
method pca/.style={color=PCAColor},
method random/.style={color=RandomColor},
adaptive diffusion/.style={solid,line width=1.25pt},
static diffusion/.style={solid,line width=1.05pt},
analytical mppca/.style={densely dotted,line width=1.20pt}
}

\tikzset{
key adasense/.style={draw=AdaBlue,solid,line width=1.25pt},
key between/.style={draw=BetweenOrange,solid,line width=1.25pt},
key fisher/.style={draw=FisherGreen,solid,line width=1.25pt},
key pca/.style={draw=PCAColor,solid,line width=1.05pt},
key random/.style={draw=RandomColor,solid,line width=1.05pt},
key mppca-full/.style={draw=AdaBlue,densely dotted,line width=1.20pt},
key mppca-between/.style={draw=BetweenOrange,densely dotted,line width=1.20pt},
key mppca-fisher/.style={draw=FisherGreen,densely dotted,line width=1.20pt}
}

\begin{figure*}[t]
\centering
\begin{tikzpicture}

\begin{groupplot}[
group style={
    group size=2 by 2,
    horizontal sep=1.6cm,
    vertical sep=0.5cm,
    group name=tradeoffsSthirtytwo
},
tradeoff axis
]

% (a) MNIST classification
\nextgroupplot[
ylabel={Accuracy (\%)},
xlabel={},
xmin=0,xmax=20,ymin=8,ymax=105,
xtick={0,5,10,15,20},
ytick={20,40,60,80,100},
extra description/.code={ \node[ anchor=north west, font=\small\bfseries ] at (rel axis cs:0.02,0.98) {(a)}; }]]
\addplot[method adasense,analytical mppca] table[x=mean_measurements,y=accuracy_percent]{mnist_class_acc_mppca_adasense.dat};
\addplot[method between,analytical mppca] table[x=mean_measurements,y=accuracy_percent]{mnist_class_acc_mppca_between.dat};
\addplot[method fisher,analytical mppca] table[x=mean_measurements,y=accuracy_percent]{mnist_class_acc_mppca_fisher.dat};
\addplot[method random,static diffusion] table[x=mean_measurements,y=accuracy_percent]{mnist_class_acc_s32_random.dat};
\addplot[method pca,static diffusion] table[x=mean_measurements,y=accuracy_percent]{mnist_class_acc_s32_pca.dat};
\addplot[method adasense,adaptive diffusion] table[x=mean_measurements,y=accuracy_percent]{mnist_class_acc_s32_adasense.dat};
\addplot[method between,adaptive diffusion] table[x=mean_measurements,y=accuracy_percent]{mnist_class_acc_s32_between.dat};
\addplot[method fisher,adaptive diffusion] table[x=mean_measurements,y=accuracy_percent]{mnist_class_acc_s32_fisher.dat};
%\addplot[method fisher,adaptive diffusion] table[x=mean_measurements,y=accuracy_percent]{generated/mnist_ridge/mnist_rf_ridges_S32_F0p0_v2.dat};
\draw[draw=InsetGray,densely dashed,line width=0.55pt] (axis cs:2,85) rectangle (axis cs:15,101);

% (b) MNIST reconstruction
\nextgroupplot[
ylabel={PSNR (dB)},
xlabel={},
xmin=0,xmax=20,ymin=11,ymax=24,
xtick={0,5,10,15,20},ytick={12,15,18,21,24},
extra description/.code={ \node[ anchor=north west, font=\small\bfseries ] at (rel axis cs:0.02,0.98) {(b)}; }]
\addplot[method adasense,analytical mppca] table[x=mean_measurements,y=mean_psnr]{mnist_class_psnr_mppca_adasense.dat};
\addplot[method between,analytical mppca] table[x=mean_measurements,y=mean_psnr]{mnist_class_psnr_mppca_between.dat};
\addplot[method fisher,analytical mppca] table[x=mean_measurements,y=mean_psnr]{mnist_class_psnr_mppca_fisher.dat};
\addplot[method random,static diffusion] table[x=mean_measurements,y=mean_psnr]{mnist_class_psnr_s32_random.dat};
\addplot[method pca,static diffusion] table[x=mean_measurements,y=mean_psnr]{mnist_class_psnr_s32_pca.dat};
\addplot[method adasense,adaptive diffusion] table[x=mean_measurements,y=mean_psnr]{mnist_class_psnr_s32_adasense.dat};
\addplot[method between,adaptive diffusion] table[x=mean_measurements,y=mean_psnr]{mnist_class_psnr_s32_between.dat};
\addplot[method fisher,adaptive diffusion] table[x=mean_measurements,y=mean_psnr]{mnist_class_psnr_s32_fisher.dat};
%\addplot[method fisher,adaptive diffusion] table[x=mean_measurements,y=mean_psnr]{generated/mnist_ridge/mnist_rf_ridges_S32_F0p0_v2.dat};

% (c) CIFAR-10 classification
\nextgroupplot[
ylabel={Accuracy (\%)},
xlabel={Mean measurements at stopping},
xmin=0,xmax=200,ymin=8,ymax=92,
xtick={0,50,100,150,200},ytick={10,30,50,70,90},
extra description/.code={ \node[ anchor=north west, font=\small\bfseries ] at (rel axis cs:0.02,0.98) {(c)}; }]]
\addplot[method adasense,analytical mppca] table[x=mean_measurements,y=accuracy_percent]{class_acc_mppca_adasense.dat};
\addplot[method between,analytical mppca] table[x=mean_measurements,y=accuracy_percent]{class_acc_mppca_between.dat};
\addplot[method fisher,analytical mppca] table[x=mean_measurements,y=accuracy_percent]{class_acc_mppca_fisher.dat};
\addplot[method random,static diffusion] table[x=mean_measurements,y=accuracy_percent]{class_acc_s32_random.dat};
\addplot[method pca,static diffusion] table[x=mean_measurements,y=accuracy_percent]{class_acc_s32_pca.dat};
\addplot[method adasense,adaptive diffusion] table[x=mean_measurements,y=accuracy_percent]{class_acc_s32_adasense.dat};
\addplot[method between,adaptive diffusion] table[x=mean_measurements,y=accuracy_percent]{class_acc_s32_between.dat};
\addplot[method fisher,adaptive diffusion] table[x=mean_measurements,y=accuracy_percent]{class_acc_s32_fisher.dat};
\draw[draw=InsetGray,densely dashed,line width=0.55pt] (axis cs:40,70) rectangle (axis cs:165,90);

% (d) CIFAR-10 reconstruction
\nextgroupplot[
ylabel={PSNR (dB)},
xlabel={Mean measurements at stopping},
xmin=0,xmax=200,ymin=12,ymax=30,
xtick={0,50,100,150,200},ytick={12,16,20,24,28},
extra description/.code={ \node[ anchor=north west, font=\small\bfseries ] at (rel axis cs:0.02,0.98) {(d)}; }]]
\addplot[method adasense,analytical mppca] table[x=mean_measurements,y=mean_psnr]{class_psnr_mppca_adasense.dat};
\addplot[method between,analytical mppca] table[x=mean_measurements,y=mean_psnr]{class_psnr_mppca_between.dat};
\addplot[method fisher,analytical mppca] table[x=mean_measurements,y=mean_psnr]{class_psnr_mppca_fisher.dat};
\addplot[method random,static diffusion] table[x=mean_measurements,y=mean_psnr]{class_psnr_s32_random.dat};
\addplot[method pca,static diffusion] table[x=mean_measurements,y=mean_psnr]{class_psnr_s32_pca.dat};
\addplot[method adasense,adaptive diffusion] table[x=mean_measurements,y=mean_psnr]{class_psnr_s32_adasense.dat};
\addplot[method between,adaptive diffusion] table[x=mean_measurements,y=mean_psnr]{class_psnr_s32_between.dat};
\addplot[method fisher,adaptive diffusion] table[x=mean_measurements,y=mean_psnr]{class_psnr_s32_fisher.dat};

\end{groupplot}

% MNIST accuracy inset
\begin{axis}[
tradeoff inset,
at={(tradeoffsSthirtytwo c1r1.north west)},anchor=north west,
xshift=3.3cm,yshift=-1.1cm,
xmin=2,xmax=15,ymin=85,ymax=101,
xtick={2,6,10,14},ytick={85,90,95,100}]
\addplot[method adasense,analytical mppca] table[x=mean_measurements,y=accuracy_percent]{mnist_class_acc_mppca_adasense.dat};
\addplot[method between,analytical mppca] table[x=mean_measurements,y=accuracy_percent]{mnist_class_acc_mppca_between.dat};
\addplot[method fisher,analytical mppca] table[x=mean_measurements,y=accuracy_percent]{mnist_class_acc_mppca_fisher.dat};
\addplot[method pca,static diffusion] table[x=mean_measurements,y=accuracy_percent]{mnist_class_acc_s32_pca.dat};
\addplot[method adasense,adaptive diffusion] table[x=mean_measurements,y=accuracy_percent]{mnist_class_acc_s32_adasense.dat};
\addplot[method between,adaptive diffusion] table[x=mean_measurements,y=accuracy_percent]{mnist_class_acc_s32_between.dat};
\addplot[method fisher,adaptive diffusion] table[x=mean_measurements,y=accuracy_percent]{mnist_class_acc_s32_fisher.dat};
%\addplot[method fisher,adaptive diffusion] table[x=mean_measurements,y=accuracy_percent]{generated/mnist_ridge/mnist_rf_ridges_S32_F0p0_v2.dat};
%\addplot[method fisher2,adaptive diffusion] table[x=mean_measurements,y=accuracy_percent]{generated/mnist_ridge/mnist_rf_ridges_S32_F1.dat};
\end{axis}

% CIFAR-10 accuracy inset
\begin{axis}[
tradeoff inset,
at={(tradeoffsSthirtytwo c1r2.north west)},anchor=north west,
xshift=3.3cm,yshift=-1.18cm,
xmin=40,xmax=165,ymin=70,ymax=90,
xtick={50,100,150},ytick={70,80,90}]
\addplot[method adasense,analytical mppca] table[x=mean_measurements,y=accuracy_percent]{class_acc_mppca_adasense.dat};
\addplot[method between,analytical mppca] table[x=mean_measurements,y=accuracy_percent]{class_acc_mppca_between.dat};
\addplot[method fisher,analytical mppca] table[x=mean_measurements,y=accuracy_percent]{class_acc_mppca_fisher.dat};
\addplot[method pca,static diffusion] table[x=mean_measurements,y=accuracy_percent]{class_acc_s32_pca.dat};
\addplot[method adasense,adaptive diffusion] table[x=mean_measurements,y=accuracy_percent]{class_acc_s32_adasense.dat};
\addplot[method between,adaptive diffusion] table[x=mean_measurements,y=accuracy_percent]{class_acc_s32_between.dat};
\addplot[method fisher,adaptive diffusion] table[x=mean_measurements,y=accuracy_percent]{class_acc_s32_fisher.dat};
\end{axis}

% One-line common legend
\node[anchor=north,inner xsep=3pt,inner ysep=1pt] at
($(tradeoffsSthirtytwo c1r2.south east)!0.5!
  (tradeoffsSthirtytwo c2r2.south west)+(0,-1cm)$)
{%
\footnotesize
\setlength{\tabcolsep}{4pt}%
\begin{tabular}{@{}ccccccccc@{}}
\tikz[baseline=-0.55ex]{\draw[key mppca-full] (0,0)--(0.6,0);} MPPCA-AS &
\tikz[baseline=-0.55ex]{\draw[key mppca-between] (0,0)--(0.6,0);} MPPCA-BC &
\tikz[baseline=-0.55ex]{\draw[key mppca-fisher] (0,0)--(0.6,0);} MPPCA-RF &
\tikz[baseline=-0.55ex]{\draw[key adasense] (0,0)--(0.6,0);} Diff-AS &
\tikz[baseline=-0.55ex]{\draw[key between] (0,0)--(0.6,0);} Diff-BC &
\tikz[baseline=-0.55ex]{\draw[key fisher] (0,0)--(0.6,0);} Diff-RF &
\tikz[baseline=-0.55ex]{\draw[key pca] (0,0)--(0.6,0);} Diff-PCA &
\tikz[baseline=-0.55ex]{\draw[key random] (0,0)--(0.6,0);} Diff-Random &
\end{tabular}%
};

\node[inner sep=0pt] at
($(tradeoffsSthirtytwo c1r2.south west)+(0,-0.9cm)$) {};

\end{tikzpicture}
\caption{Classification accuracy and reconstruction PSNR at the adaptive
stopping time on MNIST (top) and CIFAR-10 (bottom).}
\label{fig:tradeoffs_s32}
\end{figure*}
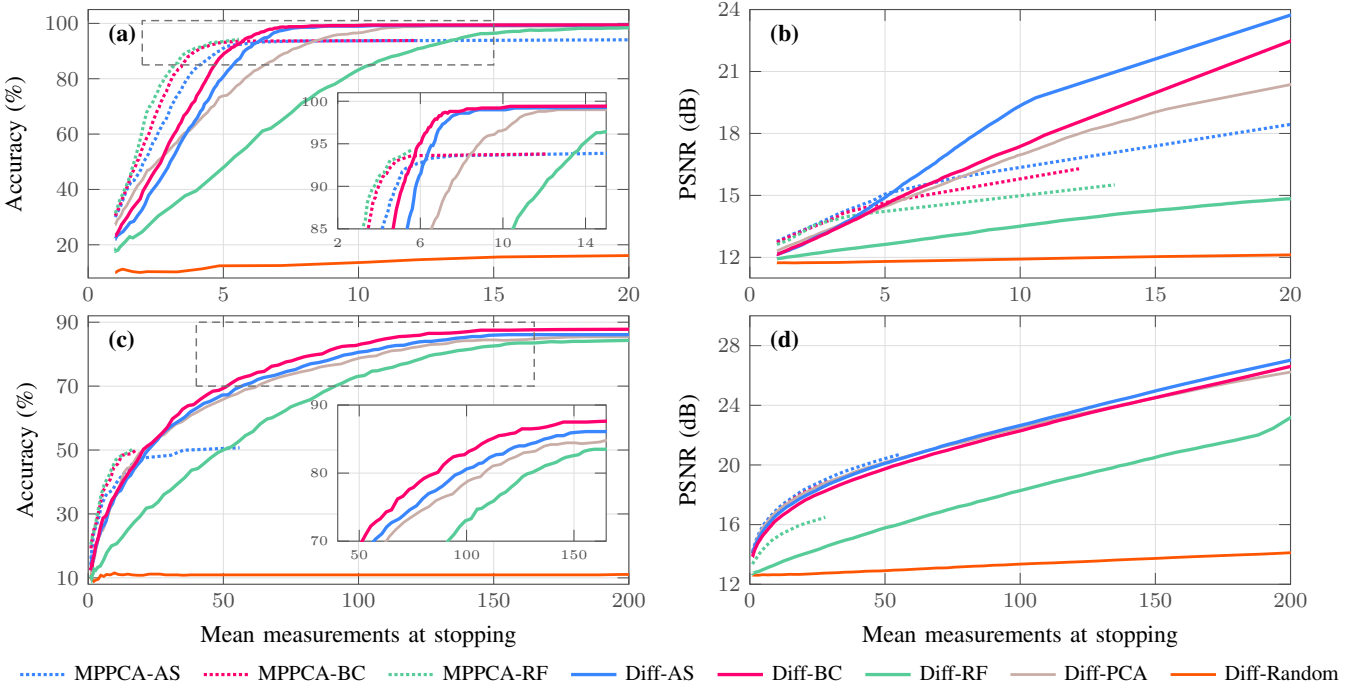

\section{Classification-Oriented Adaptive Sensing with Posterior Sampling}
\label{sec:method}
% We now map the diffusion sampling-based method to the decomposition in \eqref{eq:gmm_decomp} and to GMM quantities appearing in \eqref{eq:gmm_between_probe}--\eqref{eq:gmm_stopping_and_prediction} rules.
We now translate the GMM quantities in
\eqref{eq:gmm_decomp}--\eqref{eq:gmm_stopping_and_prediction}
to the diffusion posterior-sampling setting.
For each diffusion posterior sample in \eqref{eq:posterior_samples}, we use a
calibrated classifier to produce a vector of soft class responsibilities,
\begin{equation}
    \mathbf w_t^{(s)}
    =\softmax\!\left(\mathbf z_\phi(\vx_t^{(s)})/T\right),
    \quad
    \overline{\mathbf w}_t=\frac{1}{S}\sum_{s=1}^{S}\mathbf w_t^{(s)},
    \label{eq:diffusion_responsibilities}
\end{equation}
where $\mathbf z_\phi(\cdot)$ denotes the classifier logits and $T$ is fitted
on a disjoint calibration set by temperature scaling
~\cite{guo2017calibration}. 
% We write
% $w_{c,t}^{(s)}=[\mathbf w_t^{(s)}]_c$ and
% $\overline w_{c,t}=[\overline{\mathbf w}_t]_c$.
We use $w_{c,t}^{(s)}$ and $\overline w_{c,t}$ to denote the $c$-th
components of $\mathbf w_t^{(s)}$ and $\overline{\mathbf w}_t$, respectively.

Unlike the analytical GMM responsibility, $\overline w_{c,t}$ combines an
approximate diffusion posterior, a finite Monte Carlo sample pool, and a
learned classifier. It therefore need not equal the true posterior class
probability. Calibration is thus important when
$\max_c \overline w_{c,t}$ is used as a confidence score for stopping. Accordingly, in the diffusion-based pipeline, $\overline w_{c,t}$ replaces
$w_{c,t}$ in the stopping and prediction rules
\eqref{eq:gmm_stopping}--\eqref{eq:gmm_stopping_and_prediction}.
The threshold $\tau_{\mathrm{stop}}$ is then selected by an empirical sweep
on a disjoint validation set to achieve a target classification accuracy.
%
%\subsection{Soft Semantic Covariance from Diffusion Samples}
%\label{subsec:diffusion_semantic_covariance}

We then estimate the soft class-conditioned posterior means and between-class covariance as 
%\begin{equation}
% \widehat{\vmu}_{c,t} =\frac{\sum_s w_{c,t}^{(s)}\vx_t^{(s)}} {\sum_s w_{c,t}^{(s)}}, \qquad \widehat{\mB}_t =\sum_c \overline w_{c,t} \widehat{\vdelta}_{c,t}\widehat{\vdelta}_{c,t}^{\top}. \label{eq:soft_between} 
%\end{equation}
\begin{equation}
 \widehat{\vmu}_{c,t} =\frac{\sum_s w_{c,t}^{(s)}\vx_t^{(s)}} {\sum_s w_{c,t}^{(s)}}, \quad \widehat{\mathbf B}_t = \frac{1}{S-1} \sum_{s=1}^{S}\sum_{c=1}^{K} w_{c,t}^{(s)} \widehat{\boldsymbol{\delta}}_{c,t} \widehat{\boldsymbol{\delta}}_{c,t}^{\top}. \label{eq:soft_between} 
\end{equation}
%\begin{equation}
% \widehat{\vmu}_{c,t} =\frac{\sum_s w_{c,t}^{(s)}\vx_t^{(s)}} {\sum_s w_{c,t}^{(s)}}, \quad \widehat{\mB}_t =\sum_c \bar w_{c,t} \left( \widehat{\vmu}_{c,t} -  \widehat{\vmu}_t \right)\left( \widehat{\vmu}_{c,t} -  \widehat{\vmu}_t \right)^{\top}. \label{eq:soft_between} 
%\end{equation}
with $ \widehat{\vdelta}_{c,t} =  \widehat{\vmu}_{c,t} -  \widehat{\vmu}_t$. Thus, without assuming a Gaussian diffusion posterior, $\widehat{\mB}_t$ estimates the between-class term in \eqref{eq:gmm_decomp} from soft-labeled diffusion samples. The estimated within-class term is
%\begin{equation} \widehat{\mW}_t = \frac{1}{S} \sum_{s=1}^{S}\sum_{c=1}^{C} w_{c,t}^{(s)} \bigl(\vx_t^{(s)}-\widehat{\vmu}_{c,t}\bigr) \bigl(\vx_t^{(s)}-\widehat{\vmu}_{c,t}\bigr)^\top. \label{eq:soft_within} \end{equation}
\begin{equation} \widehat{\mW}_t = \frac{1}{S-1} \sum_{s=1}^{S}\sum_{c=1}^{K} w_{c,t}^{(s)} \bigl(\vx_t^{(s)}-\widehat{\vmu}_{c,t}\bigr) \bigl(\vx_t^{(s)}-\widehat{\vmu}_{c,t}\bigr)^\top. \label{eq:soft_within} \end{equation}
%\begin{equation} \widehat{\mS}^{\mathrm{ML}}_t = \widehat{\mW}_t+\widehat{\mB}_t. \end{equation}
%Using the same $1/S$ normalization, we can define
%\begin{equation}
%    \widehat{\mS}_t^{\mathrm{ML}}
%    =\frac{S-1}{S}\widehat{\mS}_t,
%    \qquad
%    \widehat{\mW}_t
%    =\widehat{\mS}_t^{\mathrm{ML}}-\widehat{\mB}_t,
%    \label{eq:diffusion_covariance_decomp}
%\end{equation}
%so that $\widehat{\mS}_t^{\mathrm{ML}}=\widehat{\mW}_t+\widehat{\mB}_t$ mirrors \eqref{eq:gmm_decomp}.
so that $\widehat{\mS}_t = \widehat{\mW}_t+\widehat{\mB}_t$ (defined in \eqref{eq:posterior}) mirrors \eqref{eq:gmm_decomp}.

%\subsection{Diffusion Specialization of the GMM Rules}
%\label{subsec:diffusion_classification_acquisition}

The diffusion policies follow directly from the GMM rules. Specifically,
$(\widehat{\mB}_t,\widehat{\mW}_t)$ replace $(\mB_t,\mW_t)$ in
\eqref{eq:gmm_between_probe} and \eqref{eq:gmm_fisher_probe}, while the
calibrated estimate $\overline w_{c,t}$ replaces $w_{c,t}$ in
\eqref{eq:gmm_stopping_and_prediction}. The residual-subspace constraints and
solution procedures are unchanged, and no additional optimization problem is
introduced.

%Accordingly, the three sensing strategies compared experimentally use the
%total diffusion posterior covariance $\widehat{\mS}_t$, the semantic
%between-class covariance $\widehat{\mB}_t$, or its within-class-normalized
%counterpart. Fisher normalization suppresses directions dominated by
%within-class variability, although it can become sensitive when the empirical
%within-class variance is small; $\gamma$ controls this effect.

At each stage, one posterior sample pool of size $S$ is used to estimate the class
responsibilities and covariance terms, evaluate the stopping condition, and
design the next probe. After acquisition, the posterior is resampled under
$(\mH_{t+1},\vy_{t+1})$, and all estimates are recomputed. Thus, both
the sensing direction and the signal-dependent measurement budget adapt to the
remaining posterior class ambiguity. The full approach is summarized in Fig.~\ref{fig:scheme}.

\section{Numerical evidence}

\subsection{Experimental Setup}
\label{subsec:experiments}

We evaluate the proposed framework on MNIST~\cite{lecun1998gradient} and
CIFAR-10~\cite{krizhevsky2009learning} datasets, with images scaled to $[-1,1]$. 
We instantiate the analytical GMM as a label-defined MPPCA~\cite{tipping1999mixtures}, using one low-rank PPCA component per class, with rank $r=64$ for MNIST and $r = 256$ for CIFAR-10. For diffusion posterior sampling on MNIST, we follow the AdaSense setup and use DDRM~\cite{kawar2022ddrm} inverse solver. For CIFAR-10, we use the pretrained \texttt{google/ddpm-cifar10-32} model~\cite{ho2020ddpm} with projection-conditioned reconstruction~\cite{wang2023ddnm}. In both cases, sampling is performed using DDIM~\cite{song2021ddim}, with $L=20$ reverse-diffusion steps for MNIST and $L=25$ for CIFAR-10. For MNIST, we use additive white Gaussian measurement noise with variance $\sigma_n^2=0.01$, i.e., $\mathbf{R}_t=0.01\mathbf{I}$, while the CIFAR-10 experiments are noiseless ($\mathbf{R}_t=\mathbf{0}$).  Ridge regularization $\gamma$ is set to $0.001$.

We compare diffusion-based AdaSense (Diff-AS), Between-class (Diff-BC), and regularized-Fisher (Diff-RF) acquisition using
$S=32$ posterior samples; the MPPCA variants use exact posterior
moments. All methods are evaluated on the same $1000$ test images with
$M_{\max}=32$ measurements for MNIST and $M_{\max}=256$ for CIFAR-10.
Classification uses a pretrained \texttt{resnet56} CIFAR-10 classifier and a convolutional
classifier trained on MNIST, which achieves $99.2\%$ test accuracy. Classifier probabilities are calibrated by temperature scaling~\cite{guo2017calibration},
with the temperature parameter $T$ fitted on a disjoint set of $2000$ clean images.
%We store complete acquisition trajectories
%and retrospectively sweep confidence thresholds to report classification
%accuracy and posterior-mean PSNR against the mean number of measurements at
%stopping.
%The stopping threshold and Fisher regularization are selected on
%validation data disjoint from the reported test set.

\subsection{Results and Discussion}
\label{subsec:results}

Figure~\ref{fig:tradeoffs_s32} compares AS, BC, and RF using diffusion-based and analytical MPPCA posteriors. Classification accuracy and posterior-mean reconstruction PSNR are reported as functions of the average number of measurements at stopping.

On MNIST (Fig.~\ref{fig:tradeoffs_s32}(a)-(b)), diffusion-based Diff-BC reaches approximately $95\%$ accuracy using $5.5$ measurements on average, compared with $6.5$ for Diff-AS, reducing the measurement cost by about $15\%$ (Fig.~\ref{fig:tradeoffs_s32}(a)). This advantage diminishes near saturation. 
The analytical MPPCA policies are more efficient at moderate target accuracies, reaching approximately $93\%$ accuracy with fewer measurements, but saturate below the diffusion-based policies. As expected, AS achieves the highest PSNR, while BC trades a moderate PSNR decrease for greater classification efficiency [Fig.~\ref{fig:tradeoffs_s32}(b)]. 

On CIFAR-10 (Fig.~\ref{fig:tradeoffs_s32}(c)-(d)), the MPPCA policies saturate at substantially lower accuracies, likely because a single low-rank component per class cannot capture complex class-conditional distributions. Diff-BC outperforms Diff-AS by approximately $3$--$4$ percentage points over the range of $100$--$130$ mean measurements [Fig.~\ref{fig:tradeoffs_s32}(c)]. Alternatively, reaching approximately $85\%$ accuracy requires about $110$ measurements with Diff-BC and $140$ with Diff-AS, corresponding to a $20\%$ reduction. Over this range, BC incurs a PSNR penalty of only $0.5$--$0.8$~dB [Fig.~\ref{fig:tradeoffs_s32}(d)]. 

\pgfplotsset{
posterior s8/.style={densely dashed,line width=0.95pt},
posterior s32/.style={solid,line width=1.20pt},
sample ablation column/.style={
    width=3.2cm,
    height=3cm,
    scale only axis,
    grid=major,
    major grid style={draw=GridGray,line width=0.20pt},
    axis line style={draw=AxisGray,line width=0.40pt},
    tick style={draw=AxisGray,line width=0.35pt},
    tick label style={font=\footnotesize,color=AxisGray},
    label style={font=\small},
    title style={font=\small\bfseries,yshift=-1pt},
    every axis plot/.append style={mark=none},
    unbounded coords=discard,
    clip=true
}
}

\tikzset{
key sample adasense/.style={draw=AdaBlue,solid,line width=1.20pt},
key sample between/.style={draw=BetweenOrange,solid,line width=1.20pt},
key sample fisher/.style={draw=FisherGreen,solid,line width=1.20pt},
key sample pca/.style={draw=PCAColor,solid,line width=1.20pt},
key sample s8/.style={draw=AxisGray,densely dashed,line width=0.95pt},
key sample s32/.style={draw=AxisGray,solid,line width=1.20pt}
}

\begin{figure}[t]
\centering
\begin{tikzpicture}

\begin{groupplot}[
group style={
    group size=2 by 1,
    horizontal sep=0.78cm,
    group name=sampleablationcolumn
},
sample ablation column
]

% (a) MNIST high-accuracy region
\nextgroupplot[
ylabel={Accuracy (\%)},
xmin=4,xmax=14,
ymin=90,ymax=100,
xtick={6,9,12},
ytick={90, 92, 94, 96, 98,100}, 
extra description/.code={ \node[ anchor=north west, font=\small\bfseries ] at (rel axis cs:0.01,0.99) {MNIST}; }
]
\addplot[method adasense,posterior s8]
table[x=mean_measurements,y=accuracy_percent]
{mnist_class_acc_s8_adasense.dat};
\addplot[method between,posterior s8]
table[x=mean_measurements,y=accuracy_percent]
{mnist_class_acc_s8_between.dat};
%\addplot[method fisher,posterior s8]
%table[x=mean_measurements,y=accuracy_percent]
%{mnist_class_acc_s8_fisher.dat};
\addplot[method pca,posterior s8]
table[x=mean_measurements,y=accuracy_percent]
{mnist_class_acc_s8_pca.dat};
%\addplot[method fisher, posterior s8] table[x=mean_measurements,y=accuracy_percent]{generated/mnist_ridge/mnist_rf_ridges_S8_F100_v2.dat};
\addplot[method adasense,posterior s32]
table[x=mean_measurements,y=accuracy_percent]
{mnist_class_acc_s32_adasense.dat};
\addplot[method between,posterior s32]
table[x=mean_measurements,y=accuracy_percent]
{mnist_class_acc_s32_between.dat};
%\addplot[method fisher,posterior s32]
%table[x=mean_measurements,y=accuracy_percent]
%{mnist_class_acc_s32_fisher.dat};
\addplot[method pca,posterior s32]
table[x=mean_measurements,y=accuracy_percent]
{mnist_class_acc_s32_pca.dat};
%\addplot[method fisher, posterior s32] table[x=mean_measurements,y=accuracy_percent]{generated/mnist_ridge/mnist_rf_ridges_S32_F10_v2.dat};

% (b) CIFAR-10 high-accuracy region
\nextgroupplot[
xmin=40,xmax=165,
ymin=70,ymax=90,
xtick={50,100,150},
ytick={70,75,80,85,90}, 
extra description/.code={ \node[ anchor=north west, font=\small\bfseries ] at (rel axis cs:0.01,0.99) {CIFAR-10}; }
]
\addplot[method adasense,posterior s8]
table[x=mean_measurements,y=accuracy_percent]
{class_acc_s8_adasense.dat};

%\addplot[method fisher,posterior s8]
%table[x=mean_measurements,y=accuracy_percent]
%{class_acc_s8_fisher.dat};
\addplot[method pca,posterior s8]
table[x=mean_measurements,y=accuracy_percent]
{class_acc_s8_pca.dat};
\addplot[method adasense,posterior s32]
table[x=mean_measurements,y=accuracy_percent]
{class_acc_s32_adasense.dat};
\addplot[method between,posterior s32]
table[x=mean_measurements,y=accuracy_percent]
{class_acc_s32_between.dat};
%\addplot[method fisher,posterior s32]
%table[x=mean_measurements,y=accuracy_percent]
%{class_acc_s32_fisher.dat};
\addplot[method pca,posterior s32]
table[x=mean_measurements,y=accuracy_percent]
{class_acc_s32_pca.dat};

\addplot[method between,posterior s8]
table[x=mean_measurements,y=accuracy_percent]
{class_acc_s8_between.dat};

\end{groupplot}

% Shared horizontal-axis label.
\node[anchor=north,font=\small] at
($(sampleablationcolumn c1r1.south east)!0.5!
  (sampleablationcolumn c2r1.south west)+(0,-0.3cm)$)
{Mean measurements at stopping};

% Compact factorized legend.
\node[
anchor=north,
inner xsep=3pt,
inner ysep=2pt,
fill=white
] at
($(sampleablationcolumn c1r1.south east)!0.5!
  (sampleablationcolumn c2r1.south west)+(-0.5,-0.82cm)$)
{%
\footnotesize
\setlength{\tabcolsep}{4.0pt}%
\renewcommand{\arraystretch}{1.08}%
\begin{tabular}{@{}cccccc@{}}
\tikz[baseline=-0.5ex]{\draw[key sample adasense] (0,0)--(0.4,0);} Diff-AS &
\tikz[baseline=-0.5ex]{\draw[key sample between] (0,0)--(0.4,0);} Diff-BC &
%\tikz[baseline=-0.5ex]{\draw[key sample fisher] (0,0)--(0.38,0);} RF &
\tikz[baseline=-0.5ex]{\draw[key sample pca] (0,0)--(0.4,0);} Diff-PCA &
\tikz[baseline=-0.5ex]{\draw[key sample s8] (0,0)--(0.4,0);} $S=8$ &
\tikz[baseline=-0.5ex]{\draw[key sample s32] (0,0)--(0.4,0);} $S=32$
\end{tabular}%
};

% Extend the bounding box below the legend.
\node[inner sep=0pt] at
($(sampleablationcolumn c1r1.south west)+(0,-0.9cm)$) {};

\end{tikzpicture}
\caption{Posterior sample-size ablation in the high-accuracy regions of
Fig.~\ref{fig:tradeoffs_s32}.}
\label{fig:sample_size_ablation}
\end{figure}
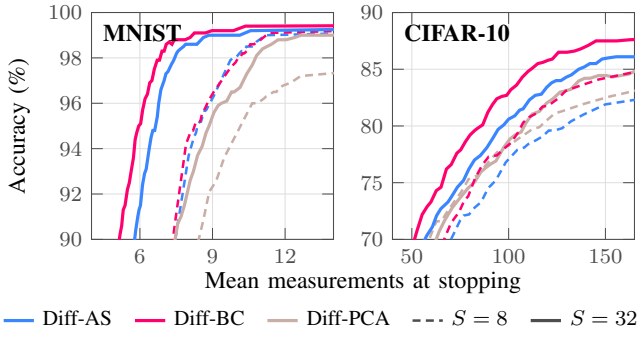

%Under the analytical MPPCA posterior, RF consistently outperforms BC. This ordering is not preserved in the diffusion setting, where RF depends on finite-sample estimates of both within-class and between-class covariance. Since $\gamma$ was not exhaustively tuned, the experiments do not establish a general ordering between BC and RF. Overall, total posterior covariance remains better suited to reconstruction, whereas between-class covariance provides a more favorable classification--measurement tradeoff.
Under the analytical MPPCA posterior, RF consistently outperforms BC. This ordering is not preserved in the diffusion setting, where RF relies on a potentially unstable finite-sample estimate of the within-class covariance. Increasing $\gamma$ reduces the influence of this estimate, and RF progressively approaches BC. Overall, total posterior covariance remains better suited to reconstruction, whereas between-class covariance provides a more favorable classification--measurement tradeoff.

To assess the benefits of statistical and image-specific adaptation, we also consider two fixed-probe baselines while keeping the rest of the pipeline unchanged. Thus, only the stopping time is image dependent. Random probes (Diff-Random) (orange curve) exploit no dataset statistics and achieve limited accuracy and PSNR. PCA probes (Diff-PCA) (gray curve) instead capture second-order statistics learned from the training set and are competitive at low measurement budgets, where adaptive probe estimation is constrained by the small posterior sample pool. Adaptive methods generally become more effective as acquisition proceeds, although Diff-PCA remains close to Diff-AS even in the high-accuracy CIFAR-10 regime. This result motivates a potential hybrid strategy that begins with a fixed batch of PCA probes before switching to image-adaptive acquisition.

Finally, Fig.~\ref{fig:sample_size_ablation} shows that reducing the posterior sample count to $S=8$ degrades the classification--measurement trade-off compared to $S=32$. On MNIST, the performance of Diff-BC approaches Diff-AS, while on CIFAR-10 BC maintains a considerable performance improvement.

\section{Conclusion}
We introduced a classification-oriented extension of diffusion-posterior adaptive sensing. A GMM analysis decomposed posterior covariance into within-class and between-class components, showing that total-variance acquisition may allocate measurements to residual variability with limited class-discriminative value. We estimated the between-class covariance from classifier-weighted diffusion posterior samples and used its dominant residual directions as adaptive probes. Experiments on MNIST and CIFAR-10 showed improved classification--measurement tradeoffs with a modest reduction in reconstruction PSNR. 
%Regularized Fisher normalization was more sensitive to finite posterior sampling, motivating further study of covariance regularization. 
Future work will consider direct expected Bayes-risk reduction, hybrid PCA/GMM--diffusion methods for faster posterior inference, and extensions to other tasks.

% IEEE requires disclosure of AI-generated content. Edit this statement to match
% the final use and the conference policy at submission time.
%\section*{Acknowledgment}
%Microsoft 365 Copilot was used to assist with language drafting and LaTeX organization. The authors reviewed, edited, and verified the technical content, equations, experiments, and conclusions.

\section*{Appendix}
\label{sec: app:gmm_derivation}
%\appendix 
%\section{Derivation of the GMM posterior decomposition} \label{app:gmm_derivation} 
Conditioned on class \(c\), the prior and measurement model are 
\[
\vx\mid c \sim \mathcal{N}(\vmu_c,\mS_c), \qquad \vy_t=\mH_t\vx+\vn_t, \qquad \vn_t\sim\mathcal{N}(\mathbf{0},\mathbf{R}_t), 
\]
where \(\vn_t\) is independent of \(\vx\). 
Therefore, 
\begin{align} \E[\vy_t\mid c] &= \mH_t\vmu_c, \quad \operatorname{Cov}(\vx,\vy_t\mid c) = \mS_c\mH_t^\top, \\ 
\operatorname{Cov}(\vy_t\mid c) &= \mH_t\mS_c\mH_t^\top+\mathbf{R}_t = \mG_{c,t}.
 \end{align} 
It follows that 
\begin{equation} 
\begin{bmatrix} \vx\\ \vy_t 
\end{bmatrix} 
\Biggm|c \sim \mathcal{N}\!\left( 
\begin{bmatrix} \vmu_c\\ \mH_t\vmu_c 
\end{bmatrix}, 
\begin{bmatrix}
 \mS_c & \mS_c\mH_t^\top \\ \mH_t\mS_c & \mG_{c,t}
 \end{bmatrix} \right).
 \label{eq:joint_class_gaussian}
 \end{equation} 
The marginal distribution of the measurement is consequently $p(\vy_t\mid c) = \mathcal{N} (\vy_t;\mH_t\vmu_c,\mG_{c,t})$.
Combining this likelihood with \(p(c)=w_c\) through Bayes' rule gives \eqref{eq:resp_main}. 
Applying the standard conditional-Gaussian formula to \eqref{eq:joint_class_gaussian} gives \eqref{eq:cond_mean_main} and \eqref{eq:cond_cov_main}. Finally, applying the law of total covariance with respect to the latent class variable gives
 \begin{align} 
\operatorname{Cov}(\vx\mid\vy_t) &= \E\!\left[ \operatorname{Cov}(\vx\mid\vy_t,c) \mid\vy_t \right] + \operatorname{Cov}\!\left( \E[\vx\mid\vy_t,c] \mid\vy_t \right). 
\label{eq:total_covariance_appendix} 
\end{align} 
Substituting $\operatorname{Cov}(\vx\mid\vy_t,c)=\mS_{c\mid t}$, $ \E[\vx\mid\vy_t,c]=\vmu_{c\mid t}$, and $p(c\mid\vy_t)=w_{c,t}$ into \eqref{eq:total_covariance_appendix} directly yields \eqref{eq:gmm_decomp}.

\bibliographystyle{IEEEbib}
\bibliography{references}

\begin{thebibliography}{10}

\bibitem{candes2006robust}
Emmanuel~J. Cand{\`e}s, Justin Romberg, and Terence Tao,
\newblock ``Robust uncertainty principles: Exact signal reconstruction from
  highly incomplete frequency information,''
\newblock {\em IEEE Transactions on Information Theory}, vol. 52, no. 2, pp.
  489--509, Feb. 2006.

\bibitem{donoho2006compressed}
David~L. Donoho,
\newblock ``Compressed sensing,''
\newblock {\em IEEE Transactions on Information Theory}, vol. 52, no. 4, pp.
  1289--1306, Apr. 2006.

\bibitem{mangia2012rakeness}
Mauro Mangia, Fabio Pareschi, Valerio Cambareri, Riccardo Rovatti, and Gianluca
  Setti,
\newblock ``Rakeness-based design of low-complexity compressed sensing,''
\newblock {\em IEEE Transactions on Circuits and Systems I: Regular Papers},
  vol. 64, no. 5, pp. 1201--1213, 2017.

\bibitem{zonzini2021model}
Federica Zonzini, Matteo Zauli, Mauro Mangia, Nicola Testoni, and Luca
  De~Marchi,
\newblock ``Model-assisted compressed sensing for vibration-based structural
  health monitoring,''
\newblock {\em IEEE Transactions on Industrial Informatics}, vol. 17, no. 11,
  pp. 7338--7347, 2021.

\bibitem{ravaglia2026adaptive}
Gabriele Ravaglia, Said Quqa, Andriy Enttsel, Mauro Mangia, Antonio Palermo,
  and Federica Zonzini,
\newblock ``Adaptive compressed sensing with masked sensing matrices and
  support estimation-based decoding for structural health monitoring
  applications,''
\newblock in {\em Proceedings of the 34th European Signal Processing Conference
  (EUSIPCO)}, Bruges, Belgium, 2026.

\bibitem{davenport2010signal}
Mark~A. Davenport, Petros~T. Boufounos, Michael~B. Wakin, and Richard~G.
  Baraniuk,
\newblock ``Signal processing with compressive measurements,''
\newblock {\em IEEE Journal of Selected Topics in Signal Processing}, vol. 4,
  no. 2, pp. 445--460, Apr. 2010.

\bibitem{davenport2007smashed}
Mark~A. Davenport, Marco~F. Duarte, Michael~B. Wakin, Jason~N. Laska, Dharmpal
  Takhar, Kevin~F. Kelly, and Richard~G. Baraniuk,
\newblock ``The smashed filter for compressive classification and target
  recognition,''
\newblock in {\em Proceedings of SPIE Computational Imaging V}, 2007, vol.
  6498, p. 64980H.

\bibitem{yu2011statistical}
Guoshen Yu and Guillermo Sapiro,
\newblock ``Statistical compressed sensing of gaussian mixture models,''
\newblock {\em IEEE Transactions on Signal Processing}, vol. 59, no. 12, pp.
  5842--5858, 2011.

\bibitem{carson2012communications}
William~R. Carson, Minhua Chen, Miguel R.~D. Rodrigues, Robert Calderbank, and
  Lawrence Carin,
\newblock ``Communications-inspired projection design with application to
  compressive sensing,''
\newblock {\em SIAM Journal on Imaging Sciences}, vol. 5, no. 4, pp.
  1185--1212, 2012.

\bibitem{duarte2013task}
Julio~M. Duarte-Carvajalino, Guoshen Yu, Lawrence Carin, and Guillermo Sapiro,
\newblock ``Task-driven adaptive statistical compressive sensing of gaussian
  mixture models,''
\newblock {\em IEEE Transactions on Signal Processing}, vol. 61, no. 3, pp.
  585--600, 2013.

\bibitem{braun2015infogreedy}
G{\'a}bor Braun, Sebastian Pokutta, and Yao Xie,
\newblock ``Info-greedy sequential adaptive compressed sensing,''
\newblock {\em IEEE Journal of Selected Topics in Signal Processing}, vol. 9,
  no. 4, pp. 601--611, June 2015.

\bibitem{schwab2019ip}
Johannes Schwab, Stephan Antholzer, and Markus Haltmeier,
\newblock ``Deep null space learning for inverse problems: convergence analysis
  and rates,''
\newblock {\em Inverse Problems}, vol. 35, no. 2, pp. 025008, jan 2019.

\bibitem{ho2020ddpm}
Jonathan Ho, Ajay Jain, and Pieter Abbeel,
\newblock ``Denoising diffusion probabilistic models,''
\newblock in {\em Advances in Neural Information Processing Systems}, 2020,
  vol.~33, pp. 6840--6851.

\bibitem{marchioni2023i2mtc}
A.~Marchioni, F.~Martinini, L.~Manovi, S.~Cortesi, R.~Rovatti, G.~Setti, and
  M.~Mangia,
\newblock ``Adapted compressed sensing with incremental encoder and deep
  performance predictor for low-power sensor node design,''
\newblock in {\em 2023 IEEE International Instrumentation and Measurement
  Technology Conference (I2MTC)}, 2023, pp. 1--6.

\bibitem{martinini2025sp}
Filippo Martinini, Mauro Mangia, Alex Marchioni, Gianluca Setti, and Riccardo
  Rovatti,
\newblock ``Incremental undersampling mri acquisition with neural self
  assessment,''
\newblock {\em Signal Processing}, vol. 228, pp. 109746, 2025.

\bibitem{elata2024adasense}
Noam Elata, Tomer Michaeli, and Michael Elad,
\newblock ``Adaptive compressed sensing with diffusion-based posterior
  sampling,''
\newblock in {\em European Conference on Computer Vision}, 2024.

\bibitem{elata2025psc}
Noam Elata, Tomer Michaeli, and Michael Elad,
\newblock ``{PSC}: Posterior sampling-based compression,''
\newblock {\em Transactions on Machine Learning Research}, 2025.

\bibitem{bingxuan2026semcom}
Bingxuan Xu, Haotian Wu, Xiaodong Xu, and Deniz Gunduz,
\newblock ``Diffusion posterior sampling with channel feedback for adaptive
  semantic communication,''
\newblock in {\em ICC 2026 - IEEE International Conference on Communications},
  2026, pp. 1--6.

\bibitem{kawar2022ddrm}
Bahjat Kawar, Michael Elad, Stefano Ermon, and Jiaming Song,
\newblock ``Denoising diffusion restoration models,''
\newblock in {\em Advances in Neural Information Processing Systems}, 2022,
  vol.~35, pp. 23593--23606.

\bibitem{fisher1936use}
Ronald~A. Fisher,
\newblock ``The use of multiple measurements in taxonomic problems,''
\newblock {\em Annals of Eugenics}, vol. 7, no. 2, pp. 179--188, 1936.

\bibitem{guo2017calibration}
Chuan Guo, Geoff Pleiss, Yu~Sun, and Kilian~Q. Weinberger,
\newblock ``On calibration of modern neural networks,''
\newblock in {\em Proceedings of the 34th International Conference on Machine
  Learning}, 2017, pp. 1321--1330.

\bibitem{lecun1998gradient}
Yann LeCun, L{\'e}on Bottou, Yoshua Bengio, and Patrick Haffner,
\newblock ``Gradient-based learning applied to document recognition,''
\newblock {\em Proceedings of the IEEE}, vol. 86, no. 11, pp. 2278--2324, 1998.

\bibitem{krizhevsky2009learning}
Alex Krizhevsky,
\newblock ``Learning multiple layers of features from tiny images,''
\newblock Tech. {R}ep., University of Toronto, 2009.

\bibitem{tipping1999mixtures}
Michael~E. Tipping and Christopher~M. Bishop,
\newblock ``Mixtures of probabilistic principal component analyzers,''
\newblock {\em Neural Computation}, vol. 11, no. 2, pp. 443--482, 1999.

\bibitem{wang2023ddnm}
Yinhuai Wang, Jiwen Yu, and Jian Zhang,
\newblock ``Zero-shot image restoration using denoising diffusion null-space
  model,''
\newblock in {\em International Conference on Learning Representations}, 2023.

\bibitem{song2021ddim}
Jiaming Song, Chenlin Meng, and Stefano Ermon,
\newblock ``Denoising diffusion implicit models,''
\newblock in {\em International Conference on Learning Representations}, 2021.

\end{thebibliography}

\end{document}